\documentclass[preprint,nofootinbib]{revtex4}

\usepackage{graphicx}

\begin{document}

\title{Evolution of density perturbations in decaying vacuum cosmology}

\author{H. A. Borges$^1$, S. Carneiro$^{1,2}$, J. C. Fabris$^3$ and C. Pigozzo$^1$}

\affiliation{$^1$Instituto de F\'{\i}sica, Universidade Federal da
Bahia, Salvador, BA, Brazil\\$^2$International Centre for
Theoretical Physics, Trieste, Italy\footnote{Associate
Member.}\\$^3$Institut d'Astrophysique de Paris, Paris,
France\footnote{Permanent address: Departamento de F\'{\i}sica,
Universidade Federal do Esp\'{\i}rito Santo, Vit\'oria, ES,
Brazil.}}

\begin{abstract}
We study cosmological perturbations in the context of an
interacting dark energy model, in which the cosmological term
decays linearly with the Hubble parameter, with concomitant matter
production. A previous joint analysis of the redshift-distance
relation for type Ia supernovas, barionic acoustic oscillations,
and the position of the first peak in the anisotropy spectrum of
the cosmic microwave background has led to acceptable values for
the cosmological parameters. Here we present our analysis of small
perturbations, under the assumption that the cosmological term,
and therefore the matter production, are strictly homogeneous.
Such a homogeneous production tends to dilute the matter contrast,
leading to a late-time suppression in the power spectrum.
Nevertheless, an excellent agreement with the observational data
can be achieved by using a higher matter density as compared to
the concordance value previously obtained. This may indicate that
our hypothesis of homogeneous matter production must be relaxed by
allowing perturbations in the interacting cosmological term.
\end{abstract}

\maketitle

\section{Introduction}

The cosmological constant problem has acquired a renewed importance
since several independent observations have been pointing to the
presence of a negative pressure component in the cosmic fluid
\cite{Pad}. From the point of view of quantum field theories, the
natural candidate for such a dark energy is the quantum vacuum.
Since, at the macroscopic level, it has the symmetry of the
background, its energy-momentum tensor has the form $T_{\mu}^{\nu}=
\Lambda g_{\mu}^{\nu}$, where $\Lambda$ is a scalar function of
coordinates. This leads, in the case of an isotropic and homogeneous
space-time and co-moving observers, to the equation of state
$p_{\Lambda} = - \rho_{\Lambda} = - \Lambda$, where $\Lambda$ may
be, in general, a function of time. In the case of a constant
$\Lambda$, the vacuum contribution plays the role of a cosmological
constant in Einstein's equations.

However, the estimation of the vacuum energy density by quantum
field theories in the flat space-time leads, after some
regularization procedure, to a very huge result when compared to the
observed value. A possible way out of this difficult is to argue
that such a result is valid only in a flat background, in which the
very Einstein equations predict a null total energy-momentum tensor.
Therefore, the huge vacuum density should be canceled by a bare
cosmological constant, like in a renormalization process. Now, if we
could obtain the vacuum density in the FLRW space-time, after the
subtraction of the Minkowskian result it would remain an effective
time-dependent $\Lambda$ term, which decreases with the expansion.

The idea of a time-dependent cosmological term has found different
phenomenological implementations \cite{Ozer}, being a subject of
renewed interest in recent years \cite{Schutzhold,Horvat,Fabris}. A
general feature of all those approaches is the production of matter,
concomitant with the vacuum decay in order to assure the covariant
conservation of the total energy \cite{Barrow}. Indeed, in the FLRW
space-time, the Bianchi identities lead to the conservation equation
\begin{equation}
\dot{\rho}_T+3H(\rho_T+p_T)=0, \label{continuidade}
\end{equation}
where $\rho_T$ and $p_T$ stand for the total energy density and
pressure, respectively, and $H = \dot{a}/a$ is the Hubble parameter.
By writing $\rho_T = \rho_m + \Lambda$ and $p_T = p_m - \Lambda$
(where $\rho_m$ and $p_m$ are the energy density and pressure of
matter), the above equation reduces to
\begin{equation}
\dot{\rho}_m + 3H(\rho_m + p_m)=-\dot{\Lambda},
\label{continuidade3}
\end{equation}
which shows that, in the case of a varying $\Lambda$, matter is not
independently conserved\footnote{Properly speaking, we should also
consider the pressure and energy associated to the very process of
matter production, that is, the energy-momentum tensor of the
interaction between matter and vacuum. In this sense, decaying
vacuum models do not differ essentially from interacting dark energy
models \cite{Zimdahl}, with the scalar function $\Lambda$ replaced
by a scalar field interacting with matter. Nevertheless, if the
vacuum decays into non-relativistic particles, as we will consider
here, the interaction term can be neglected, and the above
decomposition may be considered a good approximation.}.

An important point to be clarified in this kind of model is the
homogeneity of matter production. Of course, in a strictly
homogeneous space-time the production is homogeneous, since $\rho_m$
and $\Lambda$ depends only on time. But, in the presence of density
perturbations, is the new matter produced homogeneously, or just
where matter already exists \cite{Dirac}? In the case of a
homogeneous production, the new matter tends to dilute the density
perturbations, leading to a suppression of the density contrast. In
some models, this suppression is strong enough to impose very
restrictive observational limits to them \cite{Opher}.

In this paper we will analyze the evolution of density perturbations
in a particular, spatially flat, cosmological model with vacuum
decay \cite{Borges,Jailson}. It can be based on a phenomenological
prescription for the variation of $\Lambda$ with time \cite{SC},
given by $\Lambda \approx (H + m)^4 - m^4$, where $m$ is a
characteristic energy that can be identified with the scale of the
QCD vacuum condensation, the latest cosmological vacuum transition.
Although it can be corroborated by holographic arguments
\cite{SC,SC2}, based on the thermodynamics of de Sitter space-times,
here we will take it just as a phenomenological ansatz. In the limit
of very early times, we have $\Lambda \approx H^4$, which provides a
non-singular inflationary solution \cite{SC}.

In the opposite limit of large times we have $\Lambda = \sigma H$,
with $\sigma \approx m^3$. This scaling law for the vacuum density
was also suggested in \cite{Schutzhold}, on the basis of different
arguments. It leads to a cosmological scenario in qualitative
agreement with the standard one \cite{Borges}, with an initial
radiation era followed by a long phase dominated by dust. This dust
phase tends asymptotically to a de Sitter universe, with the
deceleration/acceleration transition occurring some time before the
present epoch. On the other hand, a quantitative analysis has shown
a good accordance with supernova observations, leading to age and
matter density parameters inside the limits imposed by other
independent observations \cite{Jailson}.

Since the radiation phase we obtain is indistinguishable from the
standard one, our analysis will be initially focused on the
evolution of density perturbations of non-relativistic matter in the
dust-dominated phase, considering wavelengths inside the
horizon\footnote{As already commented, we will assume that the
vacuum is decaying into non-relativistic particles, in order to
avoid any conflict with CMB observations and with the observed
coldness of dark matter. We will also suppose that only dark matter
is produced, since the baryon content is well constrained by
nucleosynthesis. Evidently, these assumptions cannot be verified
without a microscopic theory of the vacuum-matter interaction.}. In
this way, it will be possible to make use of a generalization of the
Newtonian linear treatment of the problem, which includes the
effects of matter production \cite{Waga}. We will show that, even in
the case of a homogeneous vacuum decay, the contrast suppression is
important only for late times, not affecting the process of galaxy
formation. On the other hand, it dominates for future times, and we
will discuss how this behavior can possibly alleviate another
problem related to the cosmological term: the cosmic coincidence
problem.

Subsequently, a relativistic analysis will be performed, in order to
construct the matter power spectrum. Again, the hypothesis of
homogeneous matter production will be used, leading as well to a
consequent power suppression. A second interesting difference as
compared to the $\Lambda$CDM model is a shift of the spectrum
turnover to the left, that is, to smaller wavenumbers. The late-time
suppression is not very sensitive to the value used for the matter
density, a feature that can already be noted from the Newtonian
analysis. On the other hand, the correction of the turnover
position, by taking a higher matter density, displaces all the
spectrum to the right, compensating the late-time power suppression.
In this way we can obtain an excellent fit of data, but with a
higher matter density in comparison with the standard case.

The article is organized as follows. In next section we review the
main features of our interacting model. In Section III we perform
the Newtonian analysis of evolution of density perturbations in the
matter era. In Section IV the matter power spectrum is constructed,
on the basis of a simplified relativistic calculation. In Section V
the reader can find our concluding remarks.

\section{The model}

The Friedmann equations in the spatially flat case are given by
(\ref{continuidade}) and $\rho_T = 3 H^2$. Let us take $\rho_T =
\rho_m + \Lambda$, $p_T = p_m - \Lambda$ and $p_m = (\gamma-1)
\rho_m$, with constant $\gamma$. Let us also take the ansatz
$\Lambda = \sigma H$, with $\sigma$ constant and positive. We obtain
the evolution equation
\begin{equation} \label{evolucao}
2\dot{H} + 3\gamma H^2 - \sigma \gamma H = 0.
\end{equation}

The solution, for $\rho_m, H > 0$, is given by \cite{Borges}
\begin{equation} \label{a}
a = C \left[\exp\left(\sigma \gamma t/2\right) -
1\right]^{\frac{2}{3\gamma}},
\end{equation}
where $a$ is the scale factor, $C$ is an integration constant, and a
second one was taken equal to zero in order to have $a = 0$ for $t =
0$.

In the radiation phase, taking $\gamma = 4/3$ and the limit of early
times ($\sigma t << 1$), we have
\begin{equation}\label{asmall}
a \approx \sqrt{2C^2\sigma t/3}.
\end{equation}
This is the same scaling law we obtain in the standard case, leading
to $H \approx 1/2t$. In the same limit we then have $\rho_m = \rho_T
- \Lambda = 3H^2 - \sigma H \approx 3H^2 = \rho_T$. By using
(\ref{asmall}) we then obtain
\begin{equation}\label{rhosmall}
\rho_T \approx \rho_m \approx \frac{\sigma^2 C^4}{3a^4} \approx
\frac{3}{4t^2},
\end{equation}
i.e., the same variation law for radiation one obtains in the
standard model, which shows that, during the radiation era, both the
cosmological term and the matter production can be dismissed.

On the other hand, in the matter era we obtain, by doing $\gamma =
1$,
\begin{equation} \label{adust}
a = C \left[\exp\left(\sigma t/2\right) - 1\right]^{\frac{2}{3}}.
\end{equation}
Taking again the limit of early times, we have
\begin{equation}\label{adustsmall}
a \approx C(\sigma t/2)^{2/3},
\end{equation}
as in the Einstein-de Sitter solution. It is also easy to see that,
in the opposite limit $t \rightarrow \infty$, (\ref{adust}) tends to
the de Sitter solution.

With the help of (\ref{adust}), and by using $\Lambda = \sigma H$
and $\rho_m = 3H^2 - \sigma H$, it is straightforward to derive the
matter and vacuum densities as functions of the scale factor. One
has
\begin{equation}\label{rhodust}
\rho_m = \frac{\sigma^2 C^3}{3a^3} + \frac{\sigma^2
C^{3/2}}{3a^{3/2}},
\end{equation}
\begin{equation}\label{Lambdadust}
\Lambda = \frac{\sigma^2}{3} + \frac{\sigma^2 C^{3/2}}{3a^{3/2}}.
\end{equation}
In these expressions, the first terms give the standard scaling of
matter (baryons included) and vacuum densities, being dominant in
the limits of early and very late times, respectively. The second
ones are owing to the process of matter production, being important
at an intermediate time scale.

With (\ref{rhodust}) and (\ref{Lambdadust}) we obtain, for the
total energy density and pressure\footnote{These are the same
expressions we obtain for a generalized Chaplygin gas
(characterized by the equation of state $p_{ch} = - A/
\rho_{ch}^{\alpha}$ \cite{Julio}), if we choose $\alpha = - 1/2$
and $A = \sqrt{\sigma^2/3}$ (see \cite{Sandro} for a detailed
discussion about this and other curious equivalences between dark
energy models). Note, however, that the oscillations in the
evolution of density perturbations characteristic of a Chaplygin
gas \cite{Waga2} are not present in our case, as we will see
below.},
\begin{equation}
\rho_T = \frac{\sigma^2}{3} \left[ \left( \frac{C}{a} \right) ^{3/2}
+ 1 \right] ^2,
\end{equation}
\begin{equation}
p_T = - \sqrt{\frac{\sigma^2}{3}} \rho_T^{1/2}.
\end{equation}

From (\ref{adust}) we can also derive the Hubble parameter as a
function of time in the matter era. It is given by
\begin{equation} \label{H}
H = \frac{\sigma/3}{1-\exp(-\sigma t/2)}.
\end{equation}
With this expression, and by using (\ref{adust}) and
(\ref{rhodust}), it is not difficult to obtain the present age of
the universe, given, in terms of the age parameter, by
\begin{equation} \label{age}
H_0 t_0 = \frac{2\ln\Omega_{m0}}{3(\Omega_{m0}-1)},
\end{equation}
where $\Omega_{m0} = \rho_{m0}/3H_0^2$ is the relative matter
density at present.

Finally, with the help of (\ref{adust}) and (\ref{H}), we can
express $H$ as a function of the redshift $z = a_0/a - 1$, which
leads to
\begin{equation} \label{Hz}
H(z) = H_0 \left[1-\Omega_{m0}+\Omega_{m0}(z+1)^{3/2}\right].
\end{equation}

With this function we have analysed the redshift-distance relation
for type Ia supernovas \cite{Jailson}, obtaining data fits as good
as with the flat $\Lambda$CDM model. With the Supernova Legacy
Survey (SNLS) \cite{SNLS} - the most confident survey we have so far
- the best fit is given by $h=0.70 \pm 0.02$ and $\Omega_{m0} = 0.32
\pm 0.05$ (with $2\sigma$), with a reduced $\chi$-square $\chi^2_r =
1.01$ (here, $h \equiv H_0/$($100$km/s.Mpc)). On the other hand, a
joint analysis of the Legacy Survey, baryonic acoustic oscillations
and the position of the first peak of CMB anisotropies has led to
the concordance values $h=0.69 \pm 0.01$ and $\Omega_{m0} = 0.36 \pm
0.01$ (with $2\sigma$), with $\chi^2_r = 1.01$ \cite{Jailson2}. With
these results one can obtain, from (\ref{age}), a universe age $t_0
\approx 15.0$ Gyr, inside the interval allowed by age estimations of
globular clusters \cite{age}.

\section{Newtonian evolution of density perturbations}

The Newtonian equation for the evolution of density perturbations in
a pressureless fluid can be generalized in order to account for
matter production \cite{Waga}. In this generalized form, it is given
by
\begin{equation} \label{Waga}
\frac{\partial^2\delta}{\partial t^2} + \left(2H +
\frac{\Psi}{\rho_m}\right) \frac{\partial\delta}{\partial t} -
\left[ \frac{\rho_m}{2} - 2H\frac{\Psi}{\rho_m} -
\frac{\partial}{\partial t}\left(\frac{\Psi}{\rho_m}\right)\right]
\delta = 0.
\end{equation}
Here, $\delta = \delta \rho_m/\rho_m$ is the density contrast of the
pressureless matter, and $\Psi$ is the source of matter production,
defined as
\begin{equation}
\dot{\rho}_m+3H(\rho_m+p_m)=\Psi. \label{continuidade2}
\end{equation}
In the case of a constant $\Lambda$, $\Psi = 0$, and (\ref{Waga})
reduces to the usual non-relativistic equation for the linear
evolution of the contrast. In our case, on the other hand, $\Psi = -
\dot{\Lambda} = - \sigma \dot{H}$, as can be seen from
(\ref{continuidade3}).

Equation (\ref{Waga}) is derived on the basis of two main
assumptions \cite{Waga}. The first one is that the produced
particles have negligible velocities as measured by observers
co-moving with the cosmic fluid. This is a reasonable hypothesis,
since we are dealing with a non-relativistic phase of universe
expansion, when $H$ (and so $\Lambda$) varies slowly enough. The
second assumption is that the vacuum component $\Lambda$ is strictly
homogeneous, which means that matter production is homogeneous as
well. This stronger hypothesis is totally ad hoc at the present
stage of the model development and, as we will see, leads to a
suppression of the contrast at large times.

In order to solve (\ref{Waga}) for our case, it is convenient to
introduce the new variable
\begin{equation} \label{x}
x=\exp(-\sigma t/2).
\end{equation}
After calculating $\rho_m$, $H$ and $\Psi$ as functions of $x$ with
the help of (\ref{adust}), (\ref{rhodust}) and (\ref{H}), equation
(\ref{Waga}) takes the form
\begin{equation} \label{contraste}
3x^2(x-1)^2\delta''+4x(x-1)\delta'-2(3x-2)\delta=0,
\end{equation}
where the prime means derivative with respect to $x$. It is possible
to show that, in the limit of early times, it reduces to the
evolution equation for the contrast in the Einstein-de Sitter model,
as should be.

The general solution of (\ref{contraste}) can be written as
\begin{equation} \label{geral}
\delta = \frac{x}{x-1} \left\{ C_1 + C_2 \left[ \frac{2}{3}
\beta(x,1/3,2/3) + x^{1/3} (x-1)^{2/3} \right] \right\},
\end{equation}
where $C_1$ and $C_2$ are integration constants to be determined by
initial conditions, and $\beta(x,a,b)$ is the incomplete beta
function, defined as
\begin{equation} \label{beta}
\beta(x,a,b)=\int_0^x\; y^{a-1}\; (1-y)^{b-1}\; dy.
\end{equation}

This $\beta$ function can be expanded in a Laurent series around
$x=1$, leading to
\begin{equation} \label{Lorrent}
\beta(x,1/3,2/3) \approx \beta(1,1/3,2/3) - \frac{3}{2}(x-1)^{2/3}.
\end{equation}
In this way, with the help of (\ref{Lorrent}) and (\ref{x}) we can
expand (\ref{geral}) around $t = 0$, obtaining
\begin{equation} \label{expansao}
\delta \approx \frac{D_1}{t} + D_2 t^{2/3},
\end{equation}
which is precisely the general solution obtained in the Einstein-de
Sitter model, as expected. The new arbitrary constants are given by
\begin{eqnarray} \label{D}
D_1 &=& -\frac{2}{\sigma} \left[ C_1 + \frac{2}{3}\beta(1,1/3,2/3)
C_2 \right],\\ D_2 &=& \frac{(4\sigma)^{2/3}}{15} C_2.
\end{eqnarray}

If, in the early time approximation (\ref{expansao}), we want to
retain just the growing mode, proportional to $t^{2/3}$, we must
choose $D_1 = 0$. Then, our general solution (\ref{geral}) reduces
to
\begin{equation} \label{solution}
\frac{\delta}{C_2} =
\frac{2\,x\;[\beta(1,1/3,2/3)-\beta(x,1/3,2/3)]}{3(1-x)} -
\frac{x^{4/3}}{(1-x)^{1/3}}.
\end{equation}
The above solution can be expressed as functions of $t$ or $a$, with
the help of (\ref{x}) and (\ref{adust}). It can also be expressed as
a function of the redshift, by using the relation
\begin{equation}
x = \frac{\Omega_{m0} (1+z)^{3/2}}{1-\Omega_{m0}+\Omega_{m0}
(1+z)^{3/2}},
\end{equation}
which can be derived with the help of (\ref{adust}), (\ref{rhodust})
and (\ref{H}).

\begin{figure}[h]
\begin{center}
\includegraphics{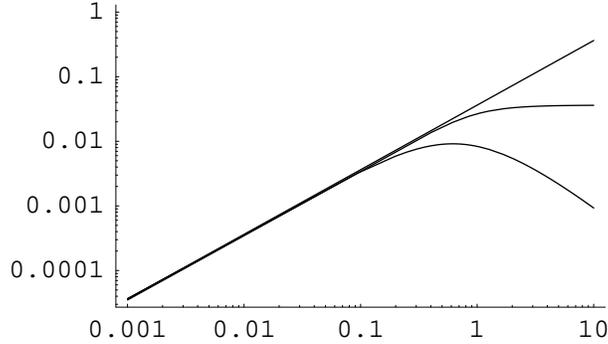}
\end{center}
\caption{\footnotesize The density contrast as a function of the
scale factor}
\end{figure}

\begin{figure}[b]
\begin{center}
\includegraphics{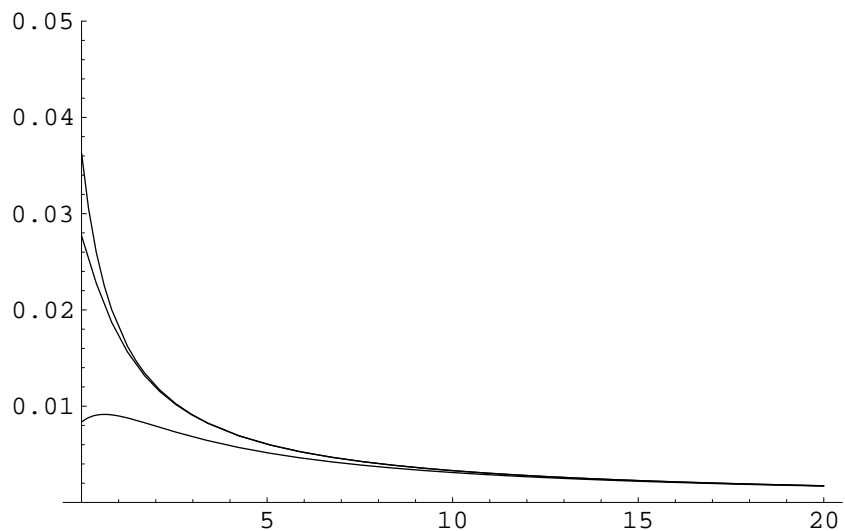}
\end{center}
\caption{\footnotesize The density contrast as a function of the
redshift}
\end{figure}

Figures 1 and 2 show the density contrast (\ref{solution}) as a
function of $a$ and $z$, respectively. We have taken $a_0 = 1$, and
used for the matter density parameter the best-fit value we have
obtained from the SNLS analysis \cite{Jailson}, $\Omega_{m0} =
0.32$. The integration constant $C_2$ was chosen so that for the
time of last scattering ($z \approx 1100$) one has $\delta \approx
10^{-5}$, as imposed by anisotropy observations of the cosmic
microwave background \cite{WMAP}. For the sake of comparison, we
have also plotted the evolution of the density contrast in the
Einstein-de Sitter solution and in the spatially flat $\Lambda$CDM
model with $\Omega_{m0} = 0.27$.

In our case the density contrast grows monotonically with time until
$z \approx 0.6$, after which it decreases monotonically, tending to
zero in the limit $t \rightarrow \infty$. The consequences of such a
suppression at large times will be discussed in our Conclusions,
where a possible relation with the cosmic coincidence problem will
be outlined. The important point here is that the evolution of
$\delta$ in our case is indistinguishable from its behavior in the
$\Lambda$CDM case until $z \approx 5$, that is, along the entire era
of galaxy formation. On the other hand, the late-time suppression
leads to a present contrast approximately $1/3$ of the standard one,
a difference that will be manifest in the power spectrum, as we will
see now.

\section{The power spectrum}

The shape of the spectrum depends on several parameters. But one of
the most important is given by the moment of equilibrium between
radiation and matter, $\Omega_R = \Omega_m$, where $\Omega_R$ and
$\Omega_m$ are the respective density parameters (relative to the
critical density). In the $\Lambda$CDM model, we have
\begin{eqnarray}
\Omega_R = \frac{\Omega_{R0}}{a^4} = \Omega_{R0}(1 + z)^4,\\
\Omega_m = \frac{\Omega_{m0}}{a^3} = \Omega_{m0}(1 + z)^3,
\end{eqnarray}
where $\Omega_{R0}$ and $\Omega_{m0}$ are the density parameters for
radiation and matter today. The redshift at equilibrium is then
given by
\begin{equation}
1 + z_{eq} = \frac{\Omega_{m0}}{\Omega_{R0}}.
\end{equation}
Following \cite{lahav}, we fix, for the $\Lambda$CDM model,
\begin{equation}
\Omega_{m0}h^2 = 0.127, \quad \Omega_{R0}h^2 = 4.1\times10^{-5}.
\end{equation}
This implies
\begin{equation}
1 + z_{eq} = 3097.
\end{equation}
Remark that this value, as a matter of fact, is independent of $h$.

Now, we can analyse the moment the perturbations enter in the
horizon. This is obtained by inspecting the perturbed equations. In
general, it can be written as
\begin{equation}
\ddot\delta + 2\frac{\dot a}{a}\dot\delta +
\biggr\{v_s^2\frac{k^2}{a^2} - \frac{3}{2}\biggr(\frac{\dot
a}{a}\biggl)^2\biggl\}\delta = 0.
\end{equation}
In this equation, $\delta$ is the density contrast, and $v_s^2 =
\frac{\partial p}{\partial\rho}$ represents the sound velocity in
unities of $c$ (the velocity of light). The presence of a first
derivative term is related to the {\it friction} due to the
expansion of the universe, while the two last terms describe the
interplay between the pressure, that avoids the collapse, and the
gravitational attraction, that drives the collapse. When the first
of these terms dominates, the perturbation does not grow; when the
second one dominates, the perturbation increases. Ignoring numerical
factors of order of unities, related to the sound velocity, equation
of state etc, the condition that separates both regimes is
\begin{equation}
k = \frac{a}{d_H}, \quad d_H = \frac{c}{H} = \frac{c\,a}{\dot a}.
\end{equation}
In this expression, $d_H$ is the Hubble radius. Of course, this is
just an estimation.

For the $\Lambda$CDM model, we have
\begin{equation}
d_H = \frac{c}{H_0}\biggr\{\Omega_{m0}(1 + z)^3 + \Omega_{R0}(1 +
z)^4 + \Omega_{\Lambda0}\biggl\}^{-1/2},
\end{equation}
where $\Omega_{\Lambda0}$ is the density parameter for the
cosmological term today. Hence, we have
\begin{equation}
[(1 + z) k\,l_{H0}]^2 = \Omega_{m0}(1 + z)^3 + \Omega_{R0}(1 + z)^4
+ \Omega_{\Lambda0},
\end{equation}
where $l_{H0}$ is the Hubble's radius today, $l_{H0} =
3000\,h^{-1}$Mpc. In general, for large values of $z$ the term
$\Omega_{\Lambda0}$ can be ignored. In doing so, and using the
expression above for $z = z_{eq}$, we find the formula (7.39) of
reference \cite{dodelson},
\begin{equation}\label{Dodelson}
k_{eq} = \sqrt{\frac{2}{\Omega_{R0}}} \frac{\Omega_{m0}}{l_{H0}}.
\end{equation}

Using, besides the values of $\Omega_{m0}$ and $\Omega_{R0}$ already
quoted, also $h = 0.7$, we obtain
\begin{equation}
k_{eq} = 0.013.
\end{equation}
We notice that, using the BBKS transfer function for the
$\Lambda$CDM model \cite{jerome}, the turning point is also located
at $k = 0.013$.

Now, the observations cover scales from $k_{min}h^{-1} = 0.010$
until $k_{max}h^{-1} = 0.185$. Using the parameters above, we find
that these modes entered in the Hubble horizon at
\begin{equation}
k_{min} \rightarrow z_1 = 2077; \quad k_{max} \rightarrow z_2 =
59143.
\end{equation}
That is, essentially, all modes entered in the radiation dominate
era.

Turning to the present interacting model, the main modifications are
the following:
\begin{enumerate}
\item The expression governing the moment the modes enter in the
Hubble horizon is given by
\begin{equation} \label{Hzalto}
\biggl [k\,l_{H0}\,(1 + z)\biggr]^2 = \frac{1}{\Omega_{m0} +
\Omega_{\Lambda0}}\biggl[\Omega_{\Lambda0} + \Omega_{m0}(1 +
z)^{3/2}\biggr]^2 + \Omega_{R0}(1 + z)^4,
\end{equation}
with $\Omega_{m0} + \Omega_{\Lambda0} \approx 1$. This is an
approximate expression obtained from (\ref{Hz}) by adding a
conserved radiation density to the Friedmann equation $3H^2 =
\rho_T$.\footnote{Note that the inclusion of conserved radiation
changes the dynamics, and, consequently, the production of matter,
$\Lambda(z)$ and $\rho_m(z)$ also change. Therefore, the exact
generalization of (\ref{Hz}) requires a reanalysis of the dynamics.
Nevertheless, as $\Omega_{R0} \approx 10^{-4} << 1$, when the vacuum
and the matter production begin to have importance, the radiation is
negligible, and vice-versa. In this way, (\ref{Hzalto}) can be
considered a very good approximation. Indeed, a numerical analysis
in the range $0 < z < 10^4$ has shown that the difference between
(\ref{Hzalto}) and the exact $H(z)$ is as small as $0.01\%$.}

\item An inspection of (\ref{Hz}) for high $z$, when $\Lambda$
and the matter production
are dismissable, shows that $\Omega_m(z) = \Omega_{m0}^2 (1+z)^3$.
In other words, we have the same scaling of conserved matter as in
the standard model, but with an extra factor $\Omega_{m0}$. This is
owing to the matter production between $t(z)$ and $t_0$: in order to
have the same matter density today, we need a smaller density at
high redshifts. As a consequence, the redshift of equilibrium
between matter and radiation is now given by $z_{eq} =
\Omega_{m0}^2/\Omega_{R0}$, while for the correspondent wave number
we obtain, instead of (\ref{Dodelson}),
\begin{equation}\label{Dodelson2}
k_{eq} = \sqrt{\frac{2}{\Omega_{R0}}} \frac{\Omega_{m0}^2}{l_{H0}}.
\end{equation}
Note the extra factor $\Omega_{m0}$ as compared to the corresponding
$\Lambda$CDM expression. As this factor is smaller than unity, this
means that the turnover of the spectrum is moved to the left, that
is, to smaller $k$'s as compared to the standard model.

\item The matter density parameter and the Hubble parameter
are not the same as before. In the subsequent analysis we will use
$\Omega_{m0} = 0.32$ and $h = 0.7$ (the type Ia supernovas best
fitting \cite{Jailson}).
\end{enumerate}

Now, the results are the following:
\begin{enumerate}
\item The equilibrium occurs at $z_{eq} = 2263$, which implies $k_{eq} = 0.007$;
\item The mode $k_{min}$ enters in the Hubble horizon at $z_1 = 3469$,
while the mode $k_{max}$ at $z_2 = 81404$.
\end{enumerate}

As already noticed, the results indicate that the spectrum is
displaced to the left, implying that there is a power suppression
with respect to the $\Lambda$CDM model. Moreover, there is, as we
have seen in the previous section, an additional power suppression
during the matter dominated phase. Hence, essentially, we must
expect that the power spectrum displays, in what concerns matter
agglomeration, an expressive power suppression in comparison with
the $\Lambda$CDM model.

However, we can displace the spectrum to the right, instead of
displace it to the left, if the values of $\Omega_{m0}$ and/or $h$
are increased. For example, for $\Omega_{m0} = 0.48$ and $h = 0.73$,
the $k_{eq}$ occurs at $0.016$, with $z_{eq} = 5094$. Moreover,
$k_{min}$ enters in the Hubble horizon at $z_1 = 2589$ and $k_{max}$
at $z_2 = 80020$. The substantial displacement to the right of
$k_{eq}$ compensates the smaller growing of perturbations during the
matter dominated phase. So, the general features of the power
spectrum are reproduced for larger values of $\Omega_{m0}$ as
compared to the $\Lambda$CDM model.

A precise derivation of the spectrum is a very though calculation,
since the Einstein-Boltzmann coupled system must be considered. A
complete analysis for the $\Lambda$CDM model leads to the so-called
BBKS transfer function \cite{jerome}, which gives the spectrum today
as function of a given primordial spectrum. For the scale invariant
spectrum, favored by the primordial inflationary scenario, the BBKS
transfer function is given by
\begin{equation}
P_m(k) = |\delta_m(k)|^2 =
AT(k)\frac{g^2(\Omega_{m0})}{g^2(\Omega_T)}k,
\end{equation}
where $A$ is a normalization of the spectrum (which can be fixed by
the spectrum of anisotropy of the cosmic microwave background
radiation), $T(k)$ is given by
\begin{eqnarray}
T(k) = \frac{\ln(1 + 2.34q)}{2.34q}\biggr[1 + 3.89q + (16.1q)^2 + (5.64q)^3
+ (6.71q)^4\biggl]^{-\frac{1}{4}}, \\
q = \frac{k}{h\Gamma}{\text Mpc}^{-1}, \quad \Gamma =
\Omega_{dm0}he^{- \Omega_{b0} - \frac{\Omega_{b0}}{\Omega_{dm0}}},
\end{eqnarray}
and where $\Omega_{m0}$, $\Omega_{dm0}$, $\Omega_{b0}$ and
$\Omega_T$ are, respectively, the present density parameters of
pressureless (baryonic + dark) matter, dark matter, baryons and the
total energy. The function $g(\Omega)$ is defined by
\begin{equation}
g(\Omega) = \frac{5}{2}\Omega\biggr[\Omega^{\frac{4}{7}} -
\Omega_{\Lambda0} + \biggr(1 + \frac{\Omega}{2}\biggl)\biggr(1 +
\frac{\Omega_{\Lambda0}}{70}\biggl)\biggl]^{- 1}.
\end{equation}
The transfer function defined above represents the fitting of the
complete numerical evaluation.

A simplified version of the transfer function, which keeps all its
essential features, can be obtained by integrating the perturbed
equations for the coupled system containing radiation and
pressureless matter, from a very high redshift until today
\cite{weinberg,winfried}. The starting point is given by the
Einstein equations and the conservation law for the energy-momentum
tensor:
\begin{eqnarray}
R_{\mu\nu} &=& 8\pi G\sum_i\biggr\{T^i_{\mu\nu}
 - \frac{1}{2}g_{\mu\nu}T^i\biggl\},\\
{T^{\mu\nu}_i}_{;\mu} &=& 0,
\end{eqnarray}
where the indice denotes the $i\,th$ fluid component. One of them
will be radiation. The other one will be the pressureless matter in
the $\Lambda$CDM case, or the vacuum-matter interacting fluids in
our case (remember that, in our case, the pressureless matter is not
independently conserved, since it interacts with vacuum).
Introducing the perturbations, $g_{\mu\nu} = g^0_{\mu\nu} +
h_{\mu\nu}$, $\rho_i = \rho_i^0 + \delta\rho$, $p_i = p_i^0 + \delta
p_i$, with ($g^0_{\mu\nu}$, $\rho_i^0$, $p_i^0$) being the
background solutions, and imposing the synchronous coordinate
condition $h_{\mu 0} = 0$, we end up with the following set of
coupled equations:
\begin{eqnarray}
\label{p1}
\ddot h + 2\frac{\dot a}{a}\dot h &=& \rho_m \delta_m + 2\rho_R \delta_R,\\
\label{p2}
\dot\delta_m - \frac{\dot\Lambda}{\rho_m}\delta_m &=& \frac{\dot h}{2},\\
\label{p3}
\dot\delta_R + \frac{4}{3}\biggr\{\frac{v}{a} - \frac{\dot h}{2}\biggl\} &=& 0,\\
\dot v &=& \frac{k^2}{4a}\delta_R,
\end{eqnarray}
where $h = h_{kk}/a^2$, $\delta_m$ and $\delta_R$ are the density
contrast for matter and radiation respectively, $v$ is connected
with the peculiar velocities of the perturbed radiative fluid, and
in the $\Lambda$CDM case $\dot{\Lambda}$ is, evidently, zero.

We now eliminate the variable $\dot h$ using (\ref{p2}), divide all
the expressions by $H_0^2$ and rewrite the resulting equations in
terms of the redshift $z$, which becomes the new dynamical variable.
In the $\Lambda$CDM case the system of equations is reduced to
\begin{eqnarray}
\delta_m'' - \frac{g[z]}{f[z]}\frac{\delta_m'}{1 + z} =
\frac{3}{2f[z]}\biggr\{\Omega_{m0}(1 + z)\delta_m
+ 2\Omega_{R0}(1 + z)^2\delta_R\biggl\},\\
\delta_R' - \frac{4}{3}\biggr\{\frac{v}{\sqrt{f[z]}} +
\delta_m'\biggl\} = 0,\\
v' = -
\biggr(\frac{k\,l_{H0}}{2}\biggl)^2\frac{\delta_R}{\sqrt{f[z]}},
\end{eqnarray}
where the primes indicate derivative with respect to the redshift
$z$. The background functions $f[z]$ and $g[z]$ are given by
\begin{eqnarray}
f[z] &=& \frac{\dot a^2}{a^2} = \Omega_{m0}(1 + z)^3 +
\Omega_{R0}(1 + z)^4 + \Omega_{\Lambda0}, \\
g[z] &=& \frac{\ddot a}{a} = - \frac{1}{2} \Omega_{m0}(1 + z)^3 -
\Omega_{R0}(1 + z)^4 + \Omega_{\Lambda0}.
\end{eqnarray}
Integrating, for example, from $z = 10^8$ (when the initial spectrum
is supposed to be scale invariant, i.e., $\delta_m, \delta_R \propto
\sqrt{k}$) until today, $z = 0$, we can reproduce the BBKS transfer
function with about $10\%$ of precision.

We can perform the same calculation for the present model, finding
the following set of perturbed equations:
\begin{eqnarray}
\delta_m'' - \biggr\{\frac{\Omega_\Lambda'}{\Omega_m} +
\frac{g_1[z]}{f_1[z]}\frac{1}{(1 + z)^2}\biggl\}\delta_m' +
\biggr\{\frac{g_1[z]}{f_1[z]}\frac{\Omega_\Lambda'}{\Omega_m}\frac{1}{(1
+ z)^2} - \frac{\Omega_\Lambda''}{\Omega_m} +
\frac{\Omega_m'\Omega_\Lambda'}{\Omega_m^2}\biggl\}\delta_m
= \nonumber\\
\frac{3}{2}\frac{1}{f_1[z](1 + z)^4}\biggr\{\Omega_m\delta_m +
2\Omega_R\delta_R\biggl\},\\
\delta_R' - \frac{4}{3}\biggr\{\frac{v}{(1 + z)\sqrt{f_1[z]}} +
\delta_m' - \frac{\Omega_\Lambda'}{\Omega_m} \delta_m \biggl\} = 0,\\
v' = - \biggl(\frac{k\,l_{H0}}{2}\biggl)^2\frac{\delta_R}{(1 +
z)\sqrt{f_1[z]}}.
\end{eqnarray}
In these equations, we use the following definitions:
\begin{eqnarray}
f_1[z] &=& \dot a^2 = \frac{1}{(\Omega_{\Lambda0} + \Omega_{m0})(1 +
z)^2}\biggr\{\Omega_{\Lambda0}
+ \Omega_{m0}(1 + z)^{3/2}\biggl\}^2\nonumber\\ &+& (1 + z)^2\Omega_{R0},\\
g_1[z] &=& \ddot a = - \frac{(1 + z)^2}{2}f_1'[z], \\
\Omega_{m}(z) &=&
\frac{\Omega_{\Lambda0}\Omega_{m0}}{\Omega_{\Lambda0} +
\Omega_{m0}}\biggr\{\frac{\Omega_{m0}}{\Omega_{\Lambda0}}(1 + z)^3 +
(1 + z)^\frac{3}{2}\biggl\}, \\
\Omega_\Lambda(z) &=& \frac{\Omega_{\Lambda0}^2}{\Omega_{\Lambda0} +
\Omega_{m0}}\biggr\{1 + \frac{\Omega_{m0}}{\Omega_{\Lambda0}}(1 +
z)^\frac{3}{2}\biggl\}.
\end{eqnarray}

\begin{figure}
\begin{center}
\includegraphics{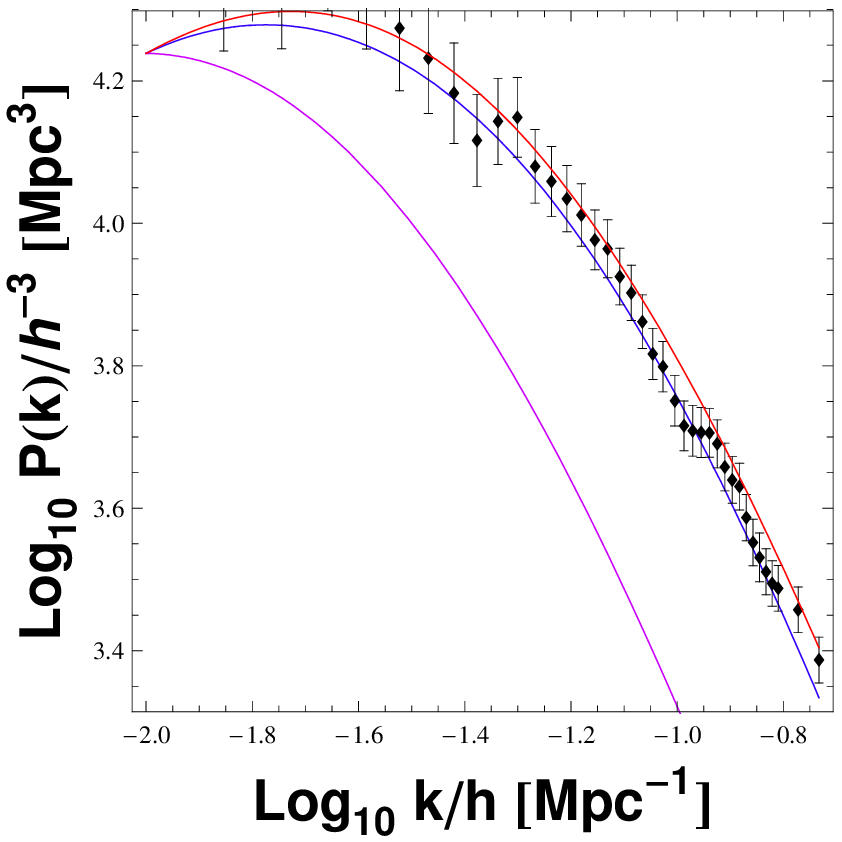}
\end{center}
\caption{\footnotesize The matter power spectra as given by the BBKS
transfer function (blue), the approximative numerical analysis used
here for $\Lambda$CDM (red) and for the interacting model (violet).
The data come from the 2dFGRS galaxy survey program \cite{2dFGRS}.
It has been used $\Omega_{m0} = 0.36$ for the interacting model.}
\end{figure}

\begin{figure}
\begin{center}
\includegraphics{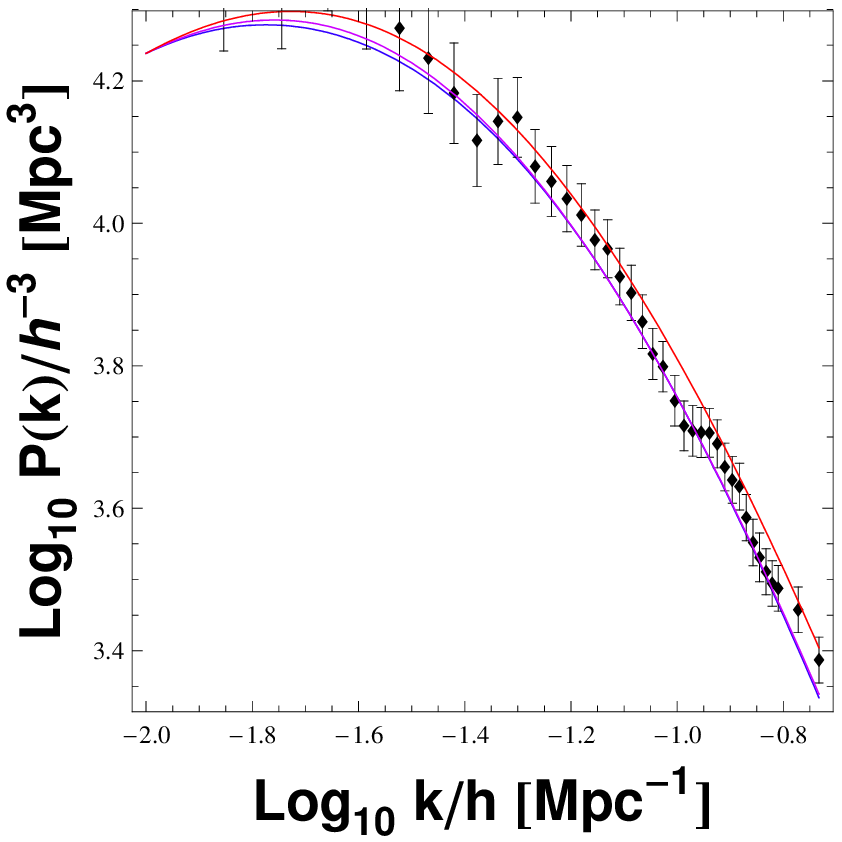}
\end{center}
\caption{\footnotesize The matter power spectra as given by the BBKS
transfer function (blue), the approximative numerical analysis used
here for $\Lambda$CDM (red) and for the interacting model (violet).
The data come from the 2dFGRS galaxy survey program \cite{2dFGRS}.
It has been used $\Omega_{m0} = 0.48$ for the interacting model.}
\end{figure}

In figures $3$ and $4$ we display the results for the exact transfer
function for the $\Lambda$CDM model (blue), the corresponding
numerical approximation (red) and the approximative transfer
function for the present model (violet). The observational data come
from the 2dFGRS galaxy survey program \cite{2dFGRS}. In the case of
the interaction model we used, in Figure $3$, $\Omega_{m0} = 0.36$,
the concordance value obtained from the joint analysis of type Ia
supernovas, BAO and CMB \cite{Jailson2}. In Figure $4$, on the other
hand, we have used $\Omega_{m0} = 0.48$. We see that in the first
case there is a substantial suppression of power, while in the
second case, where the dark matter parameter has been increased, the
agreement is excellent.

Hence, concerning the matter power spectra, the interacting model
with homogeneous matter production requires an almost double
quantity of dark matter with respect to the $\Lambda$CDM model.

\section{Conclusions}

In spite of the physical plausibility of a time dependent
cosmological term, a complete theoretical development of this idea,
including the microscopic details of the vacuum-matter interaction,
is still lacking. On the other hand, macroscopic approaches depend
on some phenomenological hypothesis, leading some times to diverse
prescriptions for the vacuum decay\footnote{For example, the linear
dependence of $\Lambda$ on the Hubble parameter we use here
contrasts with the quadratic dependence used in reference
\cite{Fabris}. In that work, the quadratic dependence is due to the
computation of quantum effects of matter field in a cosmological
background, which leads to a running cosmological term. The authors
also used the matter power spectrum data to constrain the
fundamental parameters of the quantum model.}. For this reason, a
careful comparison with current observations is very important,
playing the role of corroborating or ruling out the different
models.

We have already analysed the supernova observations \cite{Jailson},
obtaining good fits and cosmological parameters in accordance with
other independent tests, as the age of globular clusters and
dynamical limits to the matter density \cite{dynamical}. Other
precise tests, as the position of the first acoustic peak of the
cosmic microwave background and the baryonic acoustic oscillations
have also been performed \cite{Jailson2}, showing a good concordance
when jointed to the supernova analysis.

In the present paper we have studied the evolution of matter density
perturbations, in particular the contrast suppression associated to
the process of matter production. We have shown that, even in the
case of a homogeneous production, the evolution of the contrast is
the same as in the standard recipe along the entire era of galaxy
formation, diverging from the later only for $z < 5$.

On the other hand, the suppression would be dominant for future
times, and this may have an interesting relation with another
problem related to the cosmological term, namely the approximate
coincidence between the present densities of matter and dark energy.
Indeed, we can see from figures 1 and 2 that the matter contrast has
its maximum just before today, when matter and vacuum give similar
contributions to the total density. The largest structures formed
until now tend to disaggregate in the future, and their existence
then coincides with the time of approximate equality between the
matter and vacuum densities.

This could alleviate the cosmic coincidence problem, if galaxies
also follow such a process. However, we should remember that
galaxies have left the linear regime of growth a long time ago, and
that now their evolution is non-linear, driven essentially by their
self-gravitation. Therefore, an explanation of the cosmic
coincidence in the terms above will depend on a non-linear study of
density perturbations in the context of the present model. Only such
an investigation would tell us whether the contrast suppression
described here can affect smaller structures like galaxies.

Its also important to have in mind that the homogeneity of the
matter production, implicit in the derivation of solution
(\ref{solution}) and in our simplified relativistic treatment, is
just an {\it ad hoc} hypothesis, to be verified from both the
theoretical and observational viewpoints. For a constant,
non-interacting vacuum term it is certainly true, but not
necessarily in the present case. Any inhomogeneity of the vacuum
density around matter distributions may lead to an inhomogeneous
production, reducing in this way the contrast suppression. This
would allow us to fit the observed power spectrum with a smaller
matter density, closer to the concordance value obtained in
\cite{Jailson2}. Whether the matter contrast will still have a
maximum around the present time, with the discussed implications for
the coincidence problem, is a matter of investigation. A
relativistic study of this case, that is, with $\delta \Lambda \neq
0$, is already in progress.

\section*{Acknowledgements}
H. A. Borges and C. Pigozzo were supported by Capes. S. Carneiro and
J. C. Fabris are partially supported by the Brazilian Council for
Scientific Research (CNPq).

\end{document}